\documentclass[12pt]{article}
\usepackage{epsfig}

\def\BA{\begin{eqnarray}}
\def\BE{\begin{equation}}
\def\EA{\end{eqnarray}}
\def\EE{\end{equation}}

\def\Address#1#2{$^{\rm#1}${\it\footnotesize#2}\\}
\def\Ref#1{(\ref{#1})}

\def\Hs{\hat{s}}
\def\Ht{\hat{t}}
\def\Hu{\hat{u}}
\def\bb{{\bf b}}
\def\bs{{\bf s}}

\def\sNN{\sigma^{in}_{N\!N}}
\def\pp#1{p^\Vert_{#1}}
\def\pP#1{p^+_{#1}}
\def\pT#1{{\bf p}_{\bot{#1}}}
\def\abs#1{\left\vert{#1}\right\vert}
\def\dsdMdt#1{\frac{d^2\sigma{#1}}{dM^2 d\Ht}}
\def\dsdMdy#1{\frac{d^2\sigma{#1}}{dM dy}}
\def\xGmax{{x_{G\,max}}}
\def\pGmax{{p_{G\,max}}}
\def\qqG{q\!\rightarrow\!qG}
\def\qGllX{qG\!\rightarrow\!l^+\!l^-\!X}
\def\gtsim{\lower-0.45ex\hbox{$>$}\kern-0.77em\lower0.55ex\hbox{$\sim$}}

\begin{document}

% ---- Title -----------------------------------------------
\title{\bf
  Prompt Contributions to the Dilepton Yield\\
  in Heavy Ion Collisions
}

\author{
  J.~H\"ufner$^{\rm a,b}$,
  Yu.P.~Ivanov$^{\rm a,b,c}$,
  B.Z.~Kopeliovich$^{\rm b,c}$ and
  J.~Raufeisen$^{\rm a,b}$ \\
  \\
  \parbox{6.3in}{
    \Address{a}{
      Institut f\"ur Theoretische Physik der Universit\"at,
      Philosophenweg 19, D-69120 Heidelberg, Germany
    }
    \Address{b}{
      Max-Planck Institut f\"ur Kernphysik,
      Postfach 103980, D-69029 Heidelberg, Germany
    }
    \Address{c}{
      Joint Institute for Nuclear Research,
      Dubna, 141980 Moscow Region, Russia
    }
  }
}
\maketitle

% ---- Abstract --------------------------------------------
\begin{abstract}
The mass spectrum is calculated for those dileptons which are
produced in the early phase of a heavy ion collisions via the
direct production $N\!N \rightarrow l^+l^-X$ and via the
Compton process $G\!N \rightarrow l^+l^-X$ with prompt gluons
radiated in preceding $N\!N$ interactions. Both mechanisms 
produce a mass spectrum which decreases steeply with invariant
mass of the $l^+l^-$ pair and which is below the CERES data for
Pb-Au collisions by about one order of magnitude.
\end{abstract}

% ------ Introduction --------------------------------------
\section{Introduction}

Dileptons with invariant mass below 1~GeV have been measured in 
proton-nucleus \cite{pBepAu} and heavy ion collisions \cite{CERES,%
HELIOS} at the CERN SpS accelerator. For heavy ions (laboratory 
energy 158~A~GeV) a sizable enhancement of the order of a factor
three has been observed over what is calculated from the $e^+e^-$
decay of the hadrons in the final state. Also the observed shape
is very different from the calculated one. The data have not yet
found an unambiguous explanation. Most theories locate the origin
for the observed dileptons in the hot and dense phase of hadrons
\cite{Brown}-\cite{Peters}. 
This phase is rather late in the time evolution of a heavy ion
collision.

In this note, we take the opposite point of view and investigate
the mass spectrum of the dileptons which arise from the very early 
stages of the heavy ion reaction in which partons are the relevant
degrees of freedom. We calculate (i) the direct production $N\!N %
\rightarrow l^+l^-X$ in the light cone approach and in a parton 
model and (ii) the lepton production via a gluonic Compton process
$G\!N \rightarrow l^+l^-X$ from prompt gluons which are radiated in
preceding nucleon-nucleon interactions. These prompt gluons have
been recently identified as as an important source for charmonium
suppression in heavy ion collisions \cite{Charm}. This observation
has triggered the present investigation.

% ------ Direct --------------------------------------------
\section{\label{Direct}%
Direct production of dileptons: light cone approach versus parton model}

First we discuss the direct production of lepton pairs, which
is already calculated using a parton model and perturbative QCD
\cite{ApQCD}. However, in the experiments under consideration,
the invariant masses of the observed pairs are below 1~GeV. In
this domain pQCD may be questionable and we use a phenomenology
based on the light cone approach which includes nonperturbative
effects \cite{Boris,KST98,KST99}.

\begin{figure}[ht]
\centerline{\scalebox{0.8}{\includegraphics{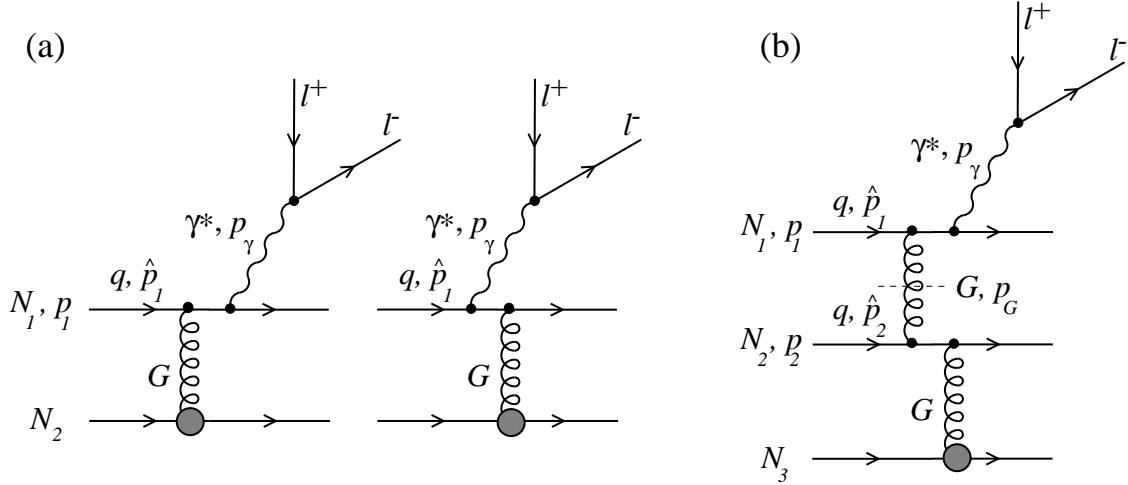}}}
\caption{
  \label{FigDiags}
  Diagrams for the two mechanisms for $l^+l^-$ production considered
  in this paper. (a) Dilepton production in $N\!N$ collisions which
  can be viewed as bremsstrahlung in $qq$ scattering (light cone
  approach) or as gluonic Compton scattering ($Gq \rightarrow %
  \gamma^*q$) from the gluon cloud of the second nucleon (parton
  approach). (b) Dileptons from prompt gluons $Gq \rightarrow %
  \gamma^*q$, where the on-mass-shell gluon G (crossed by a dashed
  line) is produced in another $N\!N$ collision.}
\end{figure}

In the light cone approach, the Drell-Yan type process illustrated by one
of the possible Feynman diagrams in Fig.~\ref{FigDiags}a is viewed in
the target rest frame, where it looks like bremsstrahlung from a $N\!N$
collision. The incident quark from $N_1$ scatters off a nucleon $N_2$
in the target and radiates a massive photon $\gamma^*$ which then decays
into a lepton pair. The cross section takes the form \cite{Boris,bhq,KST98}
\BE
  \label{formula}
  \frac{d\sigma(qN\rightarrow\gamma^* qN)}{d(\ln\alpha)} = 
    \int d^2r_T|\Psi_{\gamma^* q}(\alpha,\vec r_T)|^2
    \sigma_{q\bar q}(\alpha r_T,s),
\EE
where $\alpha$ is the fraction of the light cone momentum of the 
quark carried away by the photon and $r_T$ is the transverse separation
between the quark and the photon. $\Psi_{\gamma^* q}(\alpha,\vec r_T)$
is the light cone wave function for the transition $q \rightarrow
q\gamma^*$. We give the explicit expressions for transverse ($T$)
and longitudinal ($L$) photons
\BA
|\Psi_{\gamma^* q}(\alpha,\vec r_T)|^2   &=&
    |\Psi^T_{\gamma^* q}(\alpha,\vec r_T)|^2
  + |\Psi^L_{\gamma^* q}(\alpha,\vec r_T)|^2, \\
|\Psi^T_{\gamma^* q}(\alpha,\vec r_T)|^2 &=&
  \frac{\alpha_{em}}{\pi^2}\left\{
     m_q^2 \alpha^4 K_0^2 \left(\varepsilon r_T\right)
   + \left[1+\left(1-\alpha\right)^2\right]\varepsilon^2
     K_1^2 \left(\varepsilon r_T\right)\right\}, \\
|\Psi^L_{\gamma^* q}(\alpha,\vec r_T)|^2 &=&
  \frac{2\alpha_{em}}{\pi^2}
  M^2 \left(1-\alpha\right)^2 K_0^2 \left(\varepsilon r_T\right)
\EA
with $\varepsilon^2 = m_q^2 \alpha^2 + M^2 \left(1-\alpha\right)^2$,
where $m_q$ is the quark mass and $M^2 = p_\gamma^2$. Note that 
although there is only a single quark in the initial state the cross
section \Ref{formula} for radiation of a photon depends on the cross
section $\sigma_{q\bar q}(\alpha r_T,s)$ for the scattering of a 
$q\bar q$ dipole off a proton. Here, $\alpha r_T$ is the transverse
separation of the dipole and $s$ is the {\it c.m.} energy of the
quark nucleon system squared. 
This can be understood as follows \cite{Boris,KST98}. The physical
quark is represented as a coherent superposition of different Fock
states. In our case, we take only the bare quark $q$ and the $q\gamma^*$
states into account. If these two Fock states would scatter with the 
same amplitude off the target, coherence would be undisturbed and no
radiation produced. The radiation amplitude is proportional to the
difference between the scattering amplitudes of the two Fock components.
The dipole cross section enters, because of the interference between 
 the two graphs in fig.~(1a). The impact parameter of the projectile quark 
serves as center of gravity for the $q\gamma^*$ fluctuation. The difference in
impact parameter between the parent quark and the quark in the fluctuation is
$\alpha r_T$. Therefore, the two graphs have a relative phase factor
$\exp(i\alpha \vec r_T\cdot \vec p_T)$. 
The antiquark appears, when one takes the complex
conjugate of one of the graphs. 
It is now easy to see, that in the absolute square of the two graphs,
the color screening factor $[1-\exp(i\alpha \vec r_T\cdot \vec p_T)]$
of $\sigma_{q\bar q}(\alpha r_T,s)$ emerges
from the four different possible attachements of the two
gluons.
Therefore the dipole cross section appears in \Ref{formula}, although
there is no physical dipole in the system.

Perturbative QCD predicts $\sigma_{q\bar q}(r_T,s) \propto r_T^2$ for
$r_T\rightarrow 0$. Indeed, a colorless $q\bar q$ dipole can interact
only via its color dipole moment. However, since we are interested in
the low mass region, which corresponds to rather large separations
$r_T$, we rely on phenomenology and employ the modification \cite{KST99}
of the saturation model of \cite{Wuesthoff}
\BE
  \label{sigma}
  \sigma_{q\bar q}(r_T,s) = 
      \sigma_0(s)\left[1-\exp\left(
    - \frac{r_T^2}{r_0^2\left(s\right)}\right)\right],
\EE
where $r_0\left(s\right) = 0.88 (s/s_0)^{-0.14}~{\rm fm}$, $s_0 = %
1000~$GeV$^2$. This cross section is proportional to $r_T^2$ for 
$r_T\to 0$, but flattens off at large $r_T$. The energy dependence 
correlates with $r_T$. At small $r_T$ the dipole cross section rises
faster with energy than at large separations:
\BE
  \sigma_0(s) = 
     \sigma_{tot}^{\pi p}(s)\left(1
   + \frac{3r_0^2\left(s\right)}{8\left<r^2_{ch}\right>_{\pi}}\right),
\EE
where $\sigma_{tot}^{\pi p}(s) = 23.6(s/s_0)^{0.08}~{\rm mb}$ and
$\left<r^2_{ch}\right>_{\pi} = 0.44~{\rm fm}^2$. With this choice,
value and energy dependence of the total cross section for pion 
proton scattering is automatically reproduced. This cross section
also allows to describe well the structure function $F_2$ in DIS 
in a wide interval of energies and $Q^2$ \cite{KST99}.

The partonic cross section (\ref{formula}) has to be embedded into
the hadronic process. The cross section of direct dilepton production
takes the form
\BE
  \label{ds_D}
  \frac{d^2\sigma^{D}}{dM dy} = 
    \frac{2\alpha_{em}}{3\pi M} \sum_q e_q^2 \int_{x_1}^{1}d\alpha 
    \frac{x_1}{\alpha^2} \left[
        f_q       \left(\frac{x_1}\alpha\right)
      + f_{\bar q}\left(\frac{x_1}\alpha\right)
   \right]
   \frac{d\sigma(qN\rightarrow\gamma^* qN)}{d(\ln\alpha)}
\, +\, \Bigl\{y\Rightarrow - y\Bigr\},
\EE
where $y$ is the rapidity of the lepton pair in the c.m. frame and
$x_1=(\sqrt{x_F^2+4M^2/s}+x_F)/2$. The first term corresponds to 
radiation from the projectile quarks
and the second term to radiation from the target quarks (see Fig~%
\ref{FigDiags}a), which corresponds to replacement $y\to -y$ in the
first term. 
Since the dominant contribution to the cross section comes from large
values of $x_1/\alpha$, we neglect antiquarks in the projectile,
$f_{\bar q}\left(x_1/\alpha\right)\equiv 0$.
We parameterize the valence quark distribution in the 
well known form
\BA
  f_u(x) &=& \frac{C_u}{\sqrt x} (1-x)^3, \\
  f_d(x) &=& \frac{C_d}{\sqrt x} (1-x)^4,
\EA
where $C_{u,d}$ are defined by the normalization to the number of 
valence quarks $N_{u,d}$ in a nucleon
\BE
  \int_0^1\!\!dx \,f_{u,d}(x) = N_{u,d}.
\EE
In the case of the proton-nucleus ($pA$) and nucleus-nucleus ($AB$) 
collisions the cross section of the direct production is given
by the integration of the nuclear densities $\rho_{A,B}$ over
the impact parameter
\BE
  \label{AB_D}
  \dsdMdy{_{A\!B}^{D}} = 
    \int\!\! d\bb \int\!\! d\bs
    \int_{-\infty}^\infty\!\! dz_A\, \rho_A(\bs        ,z_A)
    \int_{-\infty}^\infty\!\! dz_B\, \rho_B(\bb\!-\!\bs,z_B)
    \, \dsdMdy{_{N\!N}}
    = AB \dsdMdy{_{N\!N}}.
\EE
In this expression we neglect small shadowing effects. To obtain
the rate of the dilepton events per interaction one has to divide
expression \Ref{AB_D} by the total cross section, which can be 
calculated using the Glauber expressions
\BA
  \sigma_{pA} &=&
    \int d\bb\left[1-\exp\left(-\sNN T_A(\bb)\right)\right], \\
  \sigma_{AB} &=&
    \int d\bb\left[1-\exp\left(-\sNN \int d\bs
    \,T_A(\bs)\,T_B(\bb-\bs)\right)\right],
\EA
where $\sNN \approx 30~{\rm mb}$ denotes the inelastic $N\!N$ cross
section and we used the standard Woods-Saxon parameterization \cite{Jager}
for the nuclear density $\rho_{A,B}$ in 
\BE
  T_{A,B}(\bb) = \int^{\infty}_{-\infty} \!\!dz\,\rho_{A,B}(\bb,z).
\EE

There are still two sources of uncertainty:
\begin{itemize}
\item The dipole cross section $\sigma_{q\bar q}(r_T,s)$ cannot be
      calculated perturbatively at large transverse separations,
      which are important in the small mass region, however. 
      We employ the phenomenological expression (\ref{sigma}).
\item The light cone wave function  $\Psi_{\gamma q}(\alpha,\vec r_T)$
      depends on the mass $m_q$ of the quark. To illustrate its influence
      we calculate the cross sections for the
      two extreme cases of the constituent quark mass $m_q = 150$~MeV
      and $m_q = 300$~MeV.
\end{itemize}

\begin{figure}[ht]
\centerline{
  \scalebox{0.5}{\includegraphics{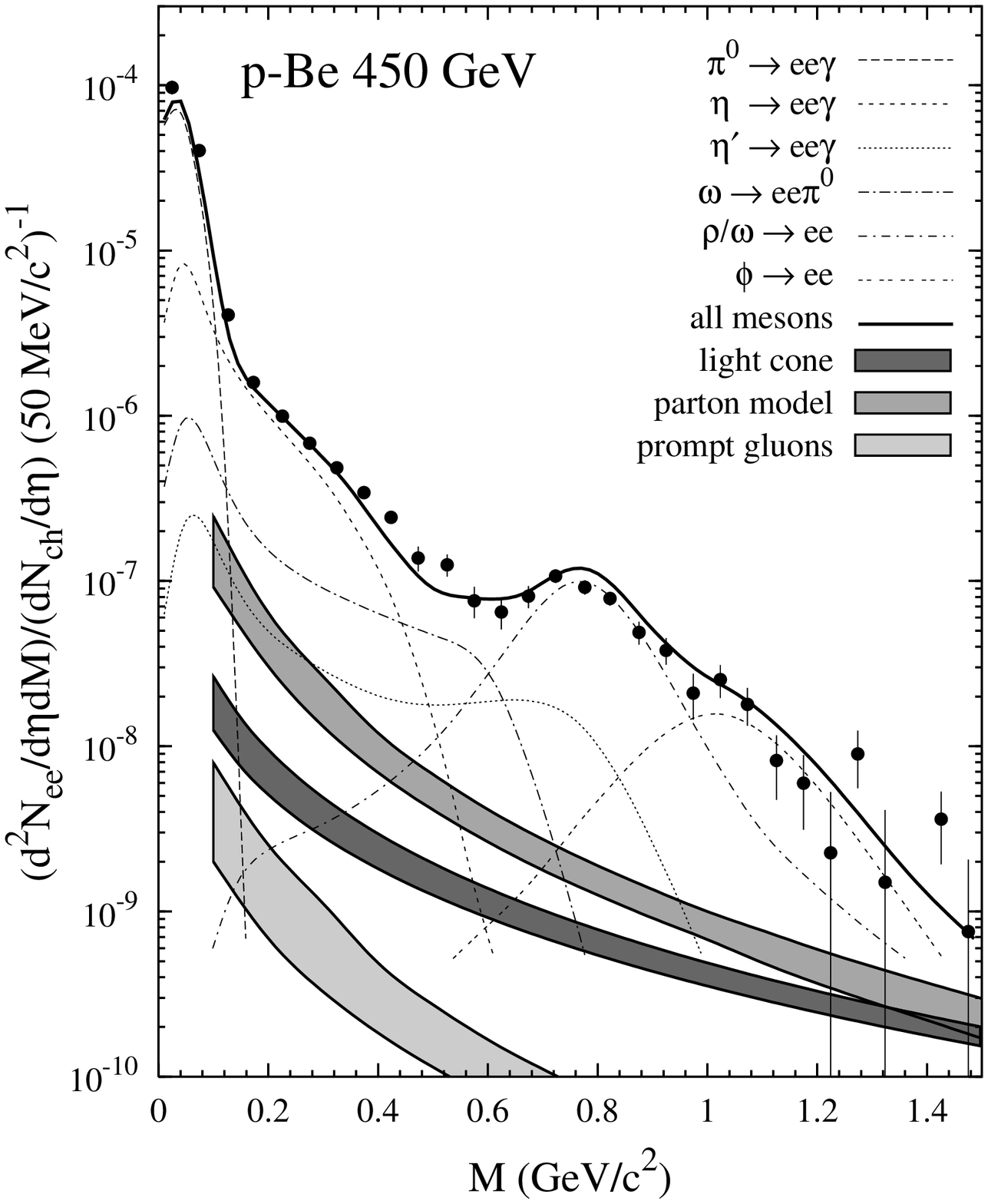}}
  \scalebox{0.5}{\includegraphics{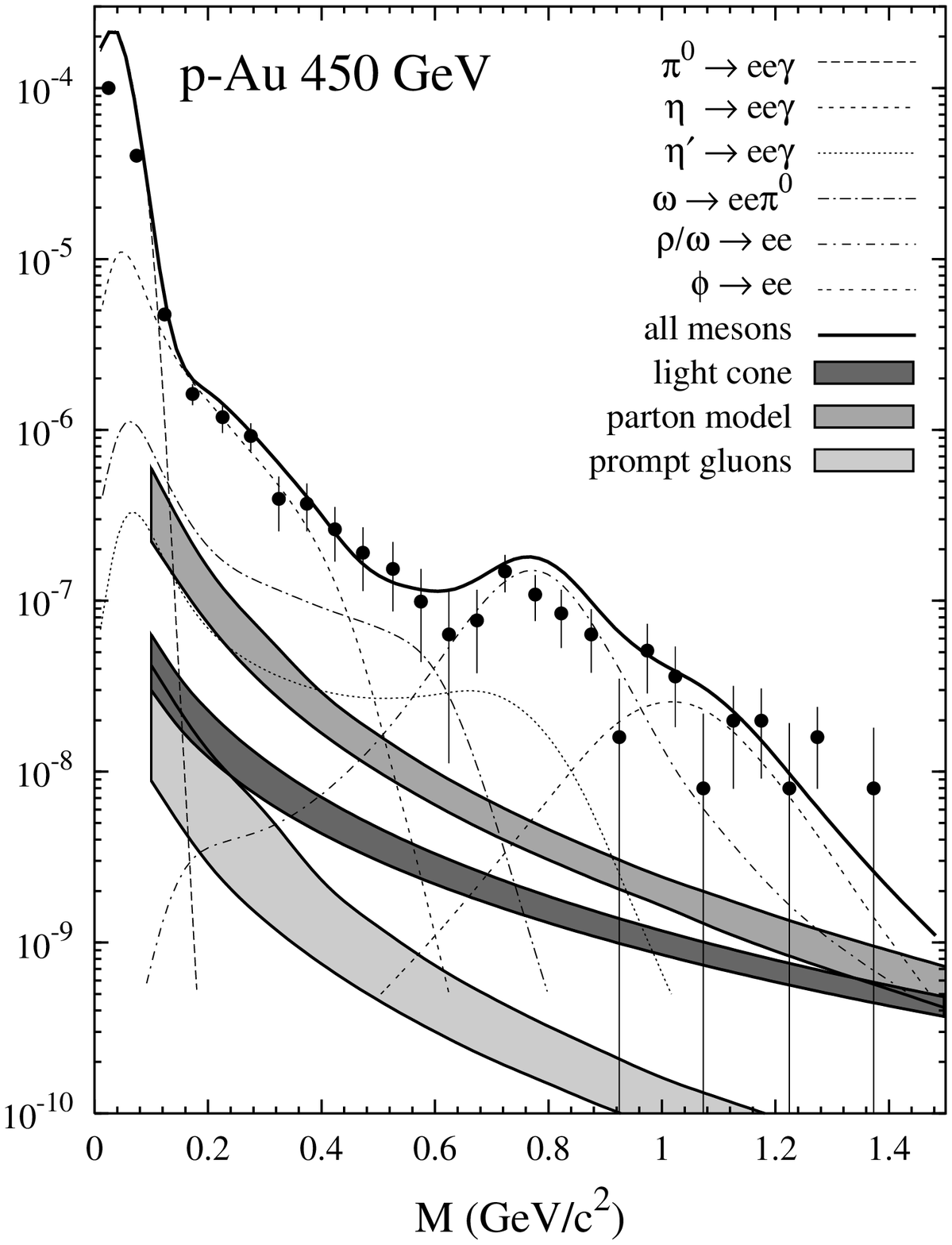}}
}
\caption{
  \label{FigBeAu}
  Data \cite{pBepAu} on dilepton events in p-Be and p-Au scattering
  together with predictions from different mechanisms. The various
  curves correspond to meson decay channels. The shaded areas represent
  our results for direct production (in the light cone and the parton
  model approaches) and via prompt gluons (see section \ref{Prompt}):
  the upper limits correspond to the quark mass $m_q=150$~MeV and the
  lower ones to $m_q=300$~MeV.
}
\end{figure}

\begin{figure}[ht]
  \scalebox{0.6}{\includegraphics{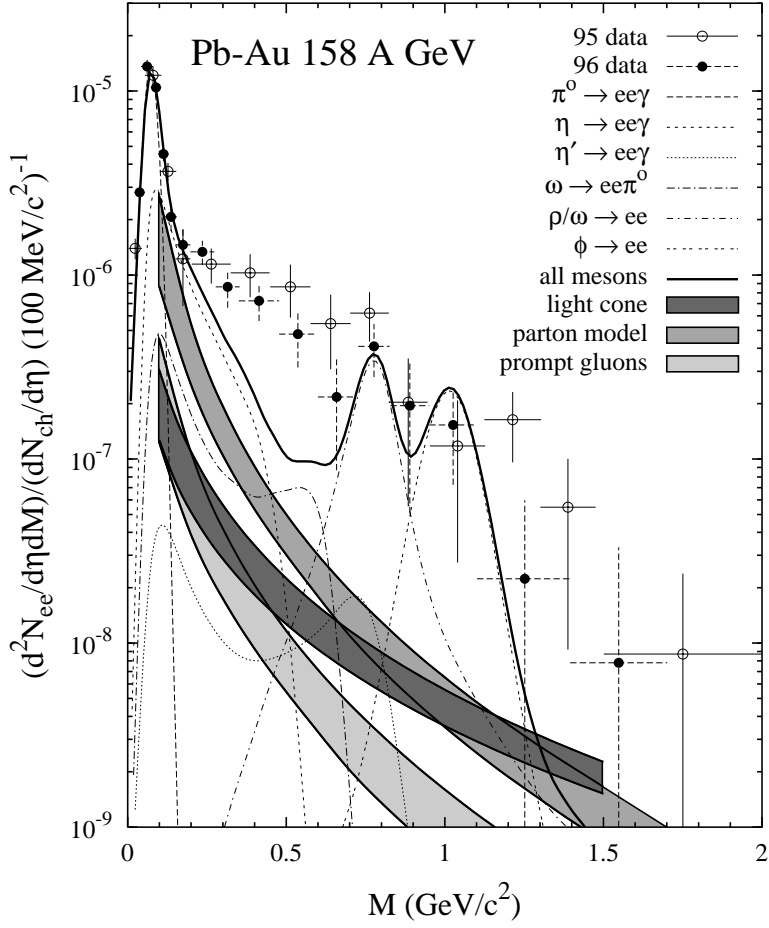}}\hfill
  \raise47mm\hbox{\parbox[b]{2.3in}{
    \caption{
      \label{FigPbAu}
      Data \cite{CERES,HELIOS} on dilepton events in Pb-Au scattering.
      The different curves correspond to meson decay channels. The 
      shaded areas are our results for direct production in the light
      cone approach and via prompt gluons (see section \ref{Prompt}):
      the upper limits correspond to the quark mass $m_q=150$~MeV and
      the lower ones to $m_q=300$~MeV. Both processes contribute
      separately to the dilepton production and therefore their
      contributions have to be added.
    }
  }
}
\end{figure}

Our results for proton (p-Be, p-Au) and heavy ion (Pb-Au) scattering
are shown in Figs.~\ref{FigBeAu},\ref{FigPbAu} under the label ``light
cone''. The border lines of the shadowed areas correspond to different 
assumptions about the constituent mass $m_q$ ($150$~MeV and $300$~MeV
for the upper and lower lines, respectively). We  believe, that also
a different dipole cross section cannot produce values, which are
significantly out of the shaded area, because $\sigma_{q\bar q}$
is well constrained to describe the structure function $F_2$ and
hadronic cross sections.

It is instructive to compare these results with predictions from
a parton model which treats direct radiation of the lepton pair
as Compton scattering of a target gluon on a beam quark (or {\em
vice versa}). This way of description should be identical to one
employed above (as one can see from Fig.~\ref{FigDiags}). However,
all phenomenological modifications introduced via the effective
$q\bar q$ cross sections are missing. The subprocess $\qGllX$ is
then calculated perturbatively with the help of the valence quark
and gluon distributions in the proton.

The main uncertainties in this case originate from:
\begin{itemize}
\item Possible failure of pQCD calculations for the soft
      Compton amplitude.
\item Poor knowledge of the gluon distribution function,
      especially at a soft scale.
\end{itemize}

We use the following gluon distribution function,
\BE
  \label{fG}
  f_G(x) = \frac{C_G}{x^{1.1}} (1-x)^5,
\EE
where $C_G$ is defined from the condition that gluons carry
a half of the nucleon momentum
\BE
  \int_0^1\!\!dx \,x \,f_G(x) = \frac12.
\EE
The pQCD calculation of the Compton amplitude is similar to
that for an interaction with on mass shell gluons presented
in the next section.

The results for proton (p-Be, p-Au) and heavy ion (Pb-Au)
scattering are shown in Figs.~\ref{FigBeAu}, \ref{FigPbAu}.
At low $M$ the results of pQCD (labeled as ``parton model'')
are higher than those from the light cone calculations but 
start to agree for $M \gtsim 1$~GeV. The reason for this
is that the saturation of the dipole cross section becomes 
important at small $M$, and the cross section decays weaker
than $1/M^3$. Although the parton model results are supposed
to coincide with the predictions of the light-cone approach
different approximations are used and the amount of disagreement
between the results may serve as a measure of theoretical 
uncertainty. Nevertheless, we believe that light cone prediction
is more reliable.

The data in figs. 2 and 3 have been measured in a pseudorapidity interval of
$2.1<\eta<2.65$, while 
our calculations have been performed for rapidity $y=2.4$.
Note that the data are also subject to $p_T$ cuts, which exclude very small
transverse momenta, $p_T<50$ MeV. 
These cuts are not included in our calculation. 
In the region of our interest ($M = 0.2$--$0.8$~GeV) however
both approaches underestimate the experimental data by at least a
factor of 10.
We therefore believe, 
that also a more careful calculation would not change our results.

% ------ Prompt --------------------------------------------
\section{\label{Prompt}%
Secondary interactions of radiated ``prompt'' gluons}

The second process, labeled ``prompt gluons'', considers gluons
which are radiated in an elementary $N\!N$ collision and which 
convert into a virtual photon via gluonic Compton scattering on
another nucleon. To estimate their contribution to the dilepton 
pair production we start with the elementary Compton subprocess
$G\,q\rightarrow\gamma^*\,q$ (see the upper part of Fig.~\ref{FigDiags}b
where the on-mass-shell gluon is marked by a dashed line). The
invariant mass distribution for $l^+l^-$ from this process can 
be expressed in terms of standard kinematic variables 
$\Hs=(\hat p_1+p_G)^2$,
$\Ht=(\hat p_1-p_\gamma)^2$ and
$\Hu=(     p_G-p_\gamma)^2$
as follows (we use expressions from \cite{ApQCD} with substitutions 
\mbox{$\Hs\Rightarrow\Hs\!-\!m_q^2$},
\mbox{$\Ht\Rightarrow\Ht\!-\!m_q^2$}, 
\mbox{$\Hu\Rightarrow\Hu\!-\!m_q^2$})
\BE
  \label{dsq_Mt}
  \dsdMdt{(\qGllX)} =
    \frac{\alpha_{em}^2 \alpha_s e^2_q}{9M^2 (\Hs-m_q^2)^2}
    \left[\frac{(\Hs-m_q^2)^2
              + (\Ht-m_q^2)^2
              + 2 M^2 (\Hu-m_q^2)}{-(\Hs-m_q^2)(\Ht-m_q^2)}
    \right].
\EE
If one neglects the transverse momenta of the initial particles, 
the cross section for the lepton pair production by the gluonic
Compton process on a quark takes the following form when going
from the variable $\Ht$ to rapidity $y$ (for notations see Fig.%
\ref{FigDiags}b)
\BE
  \label{dsq_My}
  \dsdMdy{(\qGllX)} = 2M \abs{\frac{d\Ht(y)}{dy}} \dsdMdt{(\qGllX)},
\EE
where
\BE
  \Ht(y) = m_q^2+M^2-(\Hs+M^2-m_q^2)
    \left[
      \frac{\hat E_1-\hat\pp1\tanh(y)}%
           {\hat E_1+E_G-(\hat\pp1+\pp{G})\tanh(y)}
    \right].
\EE

To obtain the cross section for a nucleon one has to integrate 
expression \Ref{dsq_My} over the quark distributions $f_q$ and 
mean number of gluons $n_G$, which are radiated by quarks from 
a projectile nucleon in an interaction with a target nucleon.
To calculate the prompt gluon spectrum we use the perturbative
evaluation for the cross section of gluon radiation \cite{KST98}
\BE
  \label{dsdadk}
  \frac{d^2\sigma(\qqG)}{d\alpha dk^2} = 
    \frac{3\alpha_s C}{\pi}
    \frac{2 m_q^2 \alpha^4 k^2
         + \left[1+(1-\alpha)^2\right]
           (k^4+\alpha^4 m_q^4)}{\left(k^2+\alpha^2 m_q^2\right)^4}
    \left[\alpha+\frac94\frac{1-\alpha}{\alpha}\right],
\EE
where $k^2 \equiv \pT{G}^2$, $C$ is the factor of the dipole
approximation for the cross section of a $q\hat q$ pair with
a nucleon (in this energy range $C \approx 3$ \cite{kz}) and
$\alpha$ is the fraction of the quark light cone momentum
carried by the gluon
\vskip -4mm
\BE
  \label{alpha}
  \alpha \equiv \frac{\pP{G}}{\hat\pP2} =
  \frac{\sqrt{\pp{G}+k^2}+\pp{G}}{\hat\pP2}.
\EE
Here light cone variables $\pP{i} \equiv E_i+\pp{i}$ are used.
By integrating \Ref{dsdadk} over $k^2$ one obtains the gluon
distribution over longitudinal momentum
\BE
  \label{nGp}
  \frac{d n_G}{d \pp{G}} = \frac{3}{\sNN}
    \int_{k^2_{min}}^{k^2_{max}} \!\!dk^2
    \abs{\frac{d\alpha}{d\pp{G}}}
    \frac{d^2\sigma(\qqG)}{d\alpha dk^2},
\EE
where
\vskip -6mm
\BA
  \label{kmax}
  k^2_{max} &=&(\hat\pP2)^2 - 2 \hat\pP2 \pp{G} \\
  \label{kmin}
  k^2_{min} &=& \max\left[ \Lambda^2,
              \eta\xi
            - \eta^2\left(\xi+m_q^2\right)
              \right],
  ~~\eta = \frac{2\hat\pP2\pp{G}+\xi}{(\hat\pP2)^2+\xi+m_q^2},
  ~~\xi  = \hat\pP2/\Delta z
\EA
and $\Lambda \approx 250$~MeV is the pQCD parameter. Expression
\Ref{kmin} is obtained from the restriction that the gluon formation
length $l_G$ should be shorter than the mean free path of a quark 
$\Delta z \approx 0.6$~fm
\vskip -4mm
\BE
  \label{lG}
  l_G = \frac{\alpha(1-\alpha)\hat\pP2}{\alpha^2m_q^2+k^2}\leq\Delta z.
\EE
For each fixed quark momentum $\hat\pp2$ eq. \Ref{nGp} gives the
gluon distribution as a function of $\pp{G}$ in the interval from
zero up to
\vskip -4mm
\BE
  \label{pGmax}
  \pGmax(\hat\pp2) = \frac{(\hat\pP2)^2-\Lambda^2}{\pP{q}}.
\EE
To obtain the prompt gluon distribution radiated by a nucleon with
momentum $p_2$ as a function of $x_G = \pp{G}/p_2$ one has to
integrate expression \Ref{nGp} with the quark distribution functions
\BA
  \label{fGprompt}
  \frac{d n_G}{d x_G} &=& \sum_q 
    \int_0^1 \!dx_2 f_q(x_2) 
    \int_0^{\pGmax(x_2 p_2)} \!\!dp_G\,\,\frac{dn_G}{dp_G}\,\,
    \delta\left(x_G-\frac{p_G}{p_2}\right) \\
           &=& p_2\,\, \sum_q 
    \int_{x_{min}}^1 \!\!dx_2\,\,f_q(x_2)\,\,
\left.\frac{dn_G}{dp_G}\right|_{p_G=x_G p_2},
\EA
where
\BE
  \label{xmin}
  x_{min} = \max\left[0,\frac{\lambda^2-m_q^2/p_2^2}{2\lambda}\right],~~~~
  \lambda = x_G+\sqrt{x_G^2+\Lambda^2/p_2^2}.
\EE
These expressions give the distributions for prompt gluons in the
interval $0 \leq x_G \leq \xGmax$
\BE
  \label{xGmax}
  \xGmax = \frac{\mu^2-\Lambda^2/p_2^2}{2\mu}, ~~~~
  \mu    = 1+\sqrt{1+m_q^2/p_2^2}.
\EE

Finally, we combine the expression \Ref{dsq_My} for the cross section
for the elementary process with the distribution function $f_q(x_1)$
for the quark and the distribution $d n_G / d x_G$ of the gluon to the
cross section for the prompt (``P'') process on the nucleon
\BE
  \label{ds_My}
  \dsdMdy{^P} = \dsdMdy{^P_+} + \dsdMdy{^P_-} ,
\EE
where
\BA
  \dsdMdy{^P_+} &=& \sum_q
     \int_0^1      \!\!dx_1 f_q(x_1) 
     \int_0^\xGmax \!\!dx_G \,\frac{d n_G}{d x_G}\,
     \dsdMdy{(\qGllX)} , \\
  \dsdMdy{^P_-} &=& \dsdMdy{^P_+}\left(y \Rightarrow -y \right) .
\EA

For nucleus-nucleus collisions the geometric factor has to take
into account that points of gluon creation and interaction are
different
\BA
  \label{AB_P}
  \dsdMdy{_{A\!B}^P} &=&
    \int\!\! d\bb \int\!\! d\bs
    \int_{-\infty}^\infty\!\! dz_A\, \rho_A(\bs        ,z_A)
    \int_{-\infty}^\infty\!\! dz_B\, \rho_B(\bb\!-\!\bs,z_B) \\
    & \times & \sNN \,
      \left[ \int_{-\infty}^{z_A} dz \rho_A(     \bs,z) \dsdMdy{^P_-}
           + \int_{z_B}^\infty dz \rho_B(\bb\!-\!\bs,z) \dsdMdy{^P_+}
      \right]. \nonumber
\EA

Again, expression \Ref{AB_P} has to be divided by the total inelastic
cross section to compare with experiment. Our results for proton (p-Be,
p-Au) and heavy ion (Pb-Au) scattering are shown in Figs.~\ref{FigBeAu},
\ref{FigPbAu}.

% ------ Conclusion ----------------------------------------
\section{Conclusion}

We have calculated two contributions to the spectrum of invariant
masses of dileptons produced in proton-nucleus and nucleus-nucleus
collisions. The direct production of dileptons $N\!N \rightarrow%
l^+l^-X$ and the production by $G\!N \rightarrow l^+l^-X$ from
gluons which have been produced in another $N\!N$ collision. Both
processes happen during the early phase of hadronization and may
be treated by pQCD on the partonic level - in contrast to the final
stage of the nuclear collision where hadrons and their decay into
$l^+l^-$ dominate.

The results of our calculations are summarized in Figs.~\ref{FigBeAu}
and \ref{FigPbAu}: the mechanisms considered contribute to the 
observed spectrum less than 10\% and therefore are unimportant on
the present level of discussion. On this level also the theoretical
uncertainties, for instance in the choice of the value for constituent
quark mass or in the evaluation via two mechanisms (light cone {\em vs.}
parton model) are not yet of importance.

Acknowledgments: We thank A.V.~Tarasov for illuminating discussions.
The work has been supported by the GSI under contract HD~H\"UF~T
and by the federal ministry BMBW under contract 06 HD 742.

% ------- Bibliography -------------------------------------

\end{document}